\documentclass[12pt]{article}
\usepackage{amssymb}

\def\be{\begin{equation}}
\def\ee{\end{equation}}
\def\bea{\begin{eqnarray}}
\def\eea{\end{eqnarray}}
\def\ba{\begin{array}}
\def\ea{\end{array}}
\def\bi{\begin{itemize}}
\def\ei{\end{itemize}}
\def\half{{\textstyle{1\over2}}}
\def\halfi{{\textstyle{i\over2}}}

\catcode`\@=11
\def\@citex[#1]#2{%
\if@filesw \immediate \write \@auxout {\string \citation {#2}}\fi
\@tempcntb\m@ne \let\@h@ld\relax \def\@citea{}%
\@cite{%
  \@for \@citeb:=#2\do {%
    \@ifundefined {b@\@citeb}%
      {\@h@ld\@citea\@tempcntb\m@ne{\bf ?}%
      \@warning {Citation `\@citeb ' on page \thepage \space undefined}}%
      {\@tempcnta\@tempcntb \advance\@tempcnta\@ne%
      \@tempcntb\number\csname b@\@citeb \endcsname \relax%
      \ifnum\@tempcnta=\@tempcntb 
        \ifx\@h@ld\relax%
          \edef \@h@ld{\@citea\csname b@\@citeb\endcsname}%
        \else%
          \edef\@h@ld{\ifmmode{-}\else--\fi\csname b@\@citeb\endcsname}%
        \fi%
      \else
        \@h@ld\@citea\csname b@\@citeb \endcsname%
        \let\@h@ld\relax%
      \fi}%
    \def\@citea{,\penalty\@highpenalty\,}%
  }\@h@ld
}{#1}}

\def\@citeb#1#2{{[#1]\if@tempswa , #2\fi}}
\def\@citeu#1#2{{$^{#1}$\if@tempswa , #2\fi }}
\def\@citep#1#2{{#1\if@tempswa , #2\fi}}

\def\bcites{         
        \catcode`\@=11
        \let\@cite=\@citeb
        \catcode`\@=12
}

\def\upcites{         
        \catcode`\@=11
        \let\@cite=\@citeu
        \catcode`\@=12
}

\def\plaincites{      
        \catcode`\@=11
        \let\@cite=\@citep
        \catcode`\@=12
}

\newcount\hour
\newcount\minute
\newtoks\amorpm
\hour=\time\divide\hour by 60
\minute=\time{\multiply\hour by 60 \global\advance\minute by-\hour}
\edef\standardtime{{\ifnum\hour<12 \global\amorpm={am}%
        \else\global\amorpm={pm}\advance\hour by-12 \fi
        \ifnum\hour=0 \hour=12 \fi
        \number\hour:\ifnum\minute<10 0\fi\number\minute\the\amorpm}}
\edef\militarytime{\number\hour:\ifnum\minute<10 0\fi\number\minute}

\def\draftlabel#1{{\@bsphack\if@filesw {\let\thepage\relax
   \xdef\@gtempa{\write\@auxout{\string
      \newlabel{#1}{{\@currentlabel}{\thepage}}}}}\@gtempa
   \if@nobreak \ifvmode\nobreak\fi\fi\fi\@esphack}
        \gdef\@eqnlabel{#1}}
\def\@eqnlabel{}
\def\@vacuum{}
\def\marginnote#1{}
\def\draftmarginnote#1{\marginpar{\raggedright\scriptsize\tt#1}}
\def\draft{
        \pagestyle{plain}
        \overfullrule=2pt
        \oddsidemargin -.5truein
        \def\@oddhead{\sl \phantom{\today\quad\militarytime} \hfil
        \smash{\Large\sl DRAFT} \hfil \today\quad\militarytime}
        \let\@evenhead\@oddhead
        \let\label=\draftlabel
        \let\marginnote=\draftmarginnote
        \def\ps@empty{\let\@mkboth\@gobbletwo
        \def\@oddfoot{\hfil \smash{\Large\sl DRAFT} \hfil}
        \let\@evenfoot\@oddhead}
        \def\@eqnnum{(\theequation)\rlap{\kern\marginparsep\tt\@eqnlabel}%
        \global\let\@eqnlabel\@vacuum}  }

\begin{document}

\hfill UTHET-02-0901

\vspace{-0.2cm}

\begin{center}
\Large
{ \bf The Penrose limit of AdS$\times S$ space and holography\footnote{Research supported by the DoE under grant DE-FG05-91ER40627.}}
\normalsize

\vspace{0.8cm}
 
{\bf George Siopsis}\footnote{gsiopsis@utk.edu}\\ Department of Physics
and Astronomy, \\
The University of Tennessee, Knoxville, \\
TN 37996 - 1200, USA.
 \end{center}

\vspace{0.8cm}
\large
\centerline{\bf Abstract}
\normalsize
\vspace{.5cm}

In the Penrose limit, AdS$\times S$ space turns into a Cahen-Wallach (CW)
space whose Killing vectors satisfy a Heisenberg algebra.
This algebra is mapped onto the holographic screen on the
boundary of AdS. I show that the Heisenberg algebra on the boundary of AdS
may be obtained directly from the CW space by appropriately constraining
the states defined on it. The transformations generated by the constraint
are similar to
gauge transformations. The ``holographic screen'' on the CW space is thus
obtained as a ``gauge-fixing'' condition.

\newpage

The Penrose limit~\cite{bib1,bibsu} of AdS$\times S$ space is obtained by boosting along a
null geodesic. If the spin in $S$ is non-vanishing, then one obtains a
Cahen-Wallach (CW)
space whose Killing vectors satisfy a Heisenberg algebra.
The sigma model (string theory) defined on such a background is exactly
solvable~\cite{bibst1,bib4a,bib4aa,bib4ab,bib4,bib4e,bib4f,bib4g,bib4h,bib4i,bib4j,bib5,bib5a,bib7,bib7a,bib2,bib3,bib3a,bib3b,bib6} and may therefore shed some light on the AdS/CFT
correspondence. Naturally, it attracted much attention resulting in an extensive
literature on the subject which cannot be adequately cited here.

The existence of an AdS/CFT
correspondence in the Penrose limit has presented a puzzle because there is no
apparent holographic screen in the CW space on which the CFT would reside,
unlike its AdS counterpart~\cite{bib9}.
A number of proposals on this issue have been made~\cite{bib8,bib8aa,bib8a,bib8b}. In the case of pure AdS space, the Penrose limit is flat Minkowski space.
This is a special case of AdS$\times S$
in which we consider a geodesic with no spin in $S$. Even though no holographic principle exists on a Minkowski space, a ``holographic screen'' can be
obtained upon restricting the states on the Minkowski space to those with a certain fixed scaling dimension.
By viewing this as a constraint on the system, the transformation it generates
can be treated similarly to a gauge transformation, even though it is not local.
The system is completely defined by data specified on a hypersurface
which cuts each orbit of ``gauge'' transformations precisely once.
This hypersurface is introduced as a ``gauge-fixing'' condition.
We thus obtain two constraints which are of {\em second class} dictating the
replacement of Poisson
brackets by Dirac brackets. This turns the Poincar\'e algebra
on the flat Minkowski space into a conformal algebra in the limit in which
the hypersurface becomes flat. The hypersurface thus defined plays the role
of the holographic screen~\cite{bibego}.

Here, I extend the discussion in~\cite{bibego} to generic null geodesics in
AdS$\times S$ space. First, I discuss how the isometries of CW space which form a Heisenberg
algebra are mapped onto the holographic screen where they act on states of large
scaling dimension $\Delta$ and $R$-charge $J$ with small (fixed) $\Delta -J$.
I then show that the holographic screen may be obtained by an appropriate restriction of the Hilbert space on CW space. This restriction generates transformations
which may be viewed as
gauge transformations in the sense described above.
Then the ``gauge-fixing'' condition is a restriction to a
hypersurface in CW space. In the limit where this hypersurface becomes flat, I
recover the expected state of affairs on the AdS holographic screen.

Let us start by fixing the notation. We are interested in the AdS$_{p+1}\times
S^q$ space.
AdS$_{p+1}$
is defined within a flat $(p+2)$-dimensional space as the hypersurface
\be\label{eqc1}
X_0^2 - X_1^2 - \dots - X_p^2 + X_{p+1}^2 = R^2
\ee
$S^q$ is the surface
\be\label{eqc1a}
Y_1^2 + \dots + Y_{q+1}^2 = R^2
\ee
in the $(q+1)$-dimensional Euclidean space $\mathbb{E}^{q+1}$.
The respective metrics in the two embedding spaces are
\be ds_\mathrm{AdS}^2 = -dX_0^2 + dX_1^2+\dots + dX_p^2 - dX_{p+1}^2\quad,\quad
ds_S^2 = dY_1^2 + \dots + dY_{q+1}^2\ee
Expressing the embedding coordinates in terms of global parameters spanning
AdS$_{p+1}$,
\be\label{eqgl} X^0 = R\cosh\rho\cos t\quad,\quad X^{p+1} = R\cosh\rho\sin t\quad,\quad
X^i = R\sinh\rho \Omega^i\quad (i=1,\dots, p)\ee
where $\Omega_1^2 + \cdots + \Omega_p^2 = 1$ span the unit sphere $S^{p-1}$,
the AdS$_{p+1}$ metric, which is inherited from the embedding upon
imposing the constraint~(\ref{eqc1}), may be written as
\be\label{eqmetric} \frac{ds_\mathrm{AdS}^2}{R^2} = -\cosh^2 \rho\, dt^2 +d\rho^2 + \sinh^2\rho\, d\Omega_{p-1}^2 \ee
Similarly, we parametrize the sphere $S^q$ by
\be Y^1 = R\cos\theta\cos\phi\quad,\quad Y^q = R\cos\theta\sin\phi\quad,\quad
Y^i = R\sin\theta \tilde\Omega^i\quad (i=2,\dots, q-1)\ee
where $\tilde\Omega_2^2 + \cdots + \tilde\Omega_{q-1}^2 = 1$ span the unit sphere $S^{q-2}$. The metric on the sphere reads
\be\label{eqmetricsph} \frac{ds_S^2}{R^2} = \cos^2 \theta\, d\phi^2 +d\theta^2 + \sin^2\theta\, d\tilde\Omega_{q-2}^2 \ee
The group of isometries of AdS$_{p+1}$ is $SO(p,2)$ whose generators we shall denote by
$K_{AB}$. In terms of the global AdS coordinates~(\ref{eqgl}), we have
\bea K_{(p+1)0} &=& i\partial_t\nonumber\\
K_{0i} &=& \cos t \left(
\Omega_i \partial_\rho + \coth\rho \nabla_i\right) -\sin t \tanh\rho \Omega_i
\partial_t\nonumber\\
K_{(p+1)i} &=& \sin t \left(
\Omega_i \partial_\rho + \coth\rho \nabla_i\right) +\cos t \tanh\rho \Omega_i
\partial_t
\eea
the rest of the generators $K_{ij}$ ($i,j = 1,\dots, p$) being the angular
momenta on the sphere $S^{p-1}$ spanned by $\Omega^i$. The algebra of the
Killing vectors is easily
verified to be $SO(p,2)$ if one uses
\be\nabla_i\Omega_j = \delta_{ij} - \Omega_i\Omega_j\ee
It is also useful to introduce the ladder operators
\be K_{\pm i} = K_{0i}\pm iK_{(p+1)i} = e^{\pm it} \left( \Omega_i (\partial_\rho \pm i\tanh\rho \partial_t) -\coth\rho \nabla_i\right)\ee
in terms of which the $SO(p,2)$ algebra reads
\be [\, K_{0(p+1)}\;,\; K_{\pm i} \, ] = \pm K_{\pm i}\quad,\quad
[\, K_{+i}\;,\; K_{-j} \, ] = 2iK_{ij} + 2\delta_{ij} K_{0(p+1)}\ee
The quadratic Casimir is
\bea\label{eqqC} \mathcal{C}_\mathrm{AdS} &=& \half K^{AB} K_{AB}\nonumber \\
&=&  K_{0(p+1)}^2 - \half \{\, K_{+i}\,,\, K_-^{\;\; i}\, \}
+ \half K_{ij}K^{ij}\nonumber \\
&=&
-\partial_\rho^2 - \frac{p\cosh^2\rho -1}{\cosh\rho\sinh\rho}\; \partial_\rho + \frac{1}{\cosh^2\rho}\, \partial_t^2 -\frac{1}{\sinh^2\rho}\;
\nabla_\Omega^2\eea
Setting $\mathcal{C}_\mathrm{AdS} = -m^2R^2$, we deduce the scalar wave equation on AdS$_{p+1}$,
\be\label{eqwe}
\left(\partial_\rho^2 + \frac{p\cosh^2\rho -1}{\cosh\rho\sinh\rho}\; \partial_\rho - \frac{1}{\cosh^2\rho}\, \partial_t^2 +\frac{1}{\sinh^2\rho}\;
\nabla_\Omega^2\right)\; \Psi^{\mathrm{(AdS)}} = m^2R^2\, \Psi^{\mathrm{(AdS)}}
\ee
Near the boundary of AdS ($\rho\to\infty$), we have $\mathcal{C}_\mathrm{AdS}
\approx \partial_\rho^2 + p\partial_\rho$. If $\Psi\sim \rho^{-\Delta}$ near
the boundary, we deduce
\be \mathcal{C}_{\mathrm{AdS}} = \Delta (\Delta -p) =m^2R^2\ee
and so
\be\Delta = \Delta_\pm = \half p \pm \half \sqrt{p^2 + 4m^2R^2}\ee
The normalizable modes have $\Delta = \Delta_+$.
The exact solution to the wave equation is
\be \Psi_{nL\vec m}^{\mathrm{(AdS)}} = e^{i\omega t}\; \tanh^L\rho\; \cosh^{-\Delta}\rho\; {}_2F_1
(-n, L+\Delta +n; L + p/2; \tanh\rho)
Y_{L\vec m} (\Omega_{p-1})\ee
where $\omega = \Delta +L + 2n$ ($n=0,1,2,\dots$), labeled by the quantum
number $n$ (or equivalently, $\omega$) and the $SO(p)$ quantum numbers
$(L,\vec m)$.
The group of isometries $SO(p,2)$ has a maximum compact subgroup $SO(2)\times
SO(p)$, where $SO(2)$ is generated by the Hamiltonian $H$, defined as the momentum conjugate to $t$ ($H = i\partial_t$),
and $SO(p)$ is the group of rotations.
The solutions to the wave equation
form a highest-weight representation of $SO(p,2)$, the highest-weight state being the one with $n=0$. It has energy $\Delta$ and
transforms trivially under the group of rotations $SO(p)$, i.e., it has $L=0$.
Explicitly,
\be \Psi_0^{\mathrm{(AdS)}} = e^{i\Delta t}\; \cosh^{-\Delta}\rho\;\ee
The solution space is built by repeatedly acting on the ground state with
the creation operators $K_{+i}$.
The boundary is conformally equivalent to an Einstein universe $S^1\times S^{p-1}$ and the $SO(p,2)$ generators turn into the generators of the conformal
group,
\be\label{eqcong} \mathcal{K}_{(p+1)0} =  i\partial_t\quad,\quad
\mathcal{K}_{\pm i} =  e^{\pm it} \left(
-\Delta\Omega_i + \nabla_i \pm i \Omega_i
\partial_t\right)
\ee
together with the $SO(p)$ generators $\mathcal{K}_{ij} = K_{ij}$.

The generators of $SO(q)$, which is the group of isometries of $S^q$, are
$J_{MN}$, where
\bea iJ_{1q} &=& \partial_\phi\nonumber\\
iJ_{1i} &=& \cos\phi \left(
\tilde\Omega_i \partial_\theta + \cot\theta \nabla_i\right) +\sin\phi \tan\theta \tilde\Omega_i
\partial_\phi\nonumber\\
iJ_{qi} &=& \sin\phi \left(
\tilde\Omega_i \partial_\theta + \cot\theta \nabla_i\right) -\cos\phi \tan\theta \tilde\Omega_i
\partial_\phi
\eea
($\tilde\Omega_2^2 +\dots +\tilde\Omega_{q-1}^2 = 1$) the rest of the generators $J_{ij}$ ($i,j = 2,\dots, q-1$) being the angular
momenta on the sphere $S^{q-2}$ spanned by $\tilde\Omega^i$.
The scalar wave equation is
\be\label{eqwes}
\frac{1}{\cos\theta\sin^{q-2}\theta}\; \frac\partial{\partial\theta} \left(\cos\theta\sin^{q-2} \frac{\partial\Psi^{(S)}}{\partial\theta}\right) + \frac{1}{\cos^2\theta}\, \partial_\phi^2\, \Psi^{(S)} +\frac{1}{\sin^2\theta}\,\nabla_{\tilde\Omega_{q-2}}^2 \Psi^{(S)} = -\mathcal{C}_S\, \Psi^{(S)}
\ee
where $\mathcal{C}_S = J(J+q)$ is the quadratic Casimir for the sphere $S^q$,
\be\label{eqqCsph} \mathcal{C}_S = \half J^{MN} J_{MN}\ee
The highest-weight state is the eigenfunction
of the ``Hamiltonian'' $-i\partial_\phi$ corresponding to the lowest eigenvalue
$-J$,
\be\Psi_0^{(S)} = e^{-iJ\phi}\cos^J\theta\ee
The other states are obtained by acting with the ladder operators
\be J_{+i} = J_{1i}+iJ_{qi} = i\, e^{i\phi} \left( \tilde\Omega_i (\partial_\theta - i\tan\theta \partial_\phi) +\cot\theta \nabla_i\right)\ee
The Penrose limit can be obtained as the scaling limit $R\to\infty$ and $\rho = r/R$, $t= t_- - t_+/R^2$,
$\theta = u/R^2$, $\phi = t_- + t_+/R^2$.
The metric in AdS space~(\ref{eqmetric}) turns into
\be ds_{AdS}^2 = -(R^2+ r^2) dt^2 + dr^2 +r^2 d\Omega_{p-1}^2 +o(1/R^2)\ee
and the sphere~(\ref{eqmetricsph}) goes to
\be\label{eqm2} ds_S^2 = (R^2- u^2) d\phi^2 + du^2 +u^2 d\tilde\Omega_{q-2}^2 +o(1/R^2)\ee
The combined space is Cahen-Wallach space in Brinkman form,
\be\label{eqm3} ds_{AdS}^2 + ds_S^2 \rightarrow ds_{CW}^2 = 4dt^+dt^- -(r^2+u^2)(dt^-)^2 + dr^2 +r^2 d\Omega_{p-1}^2 + du^2 +u^2 d\tilde\Omega_{q-2}^2\ee
The Killing vectors turn into
\bea\label{eqcong2} \frac{i}{R^2}\, K_{(p+1)0} &\rightarrow\quad e_+ &=\quad \half\partial_+\nonumber\\
\frac{1}{R}\, K_{0i} &\rightarrow\quad e_i^\ast &=\quad \cos t^- \partial_{x^i} +\half\sin t^- x_i
\partial_+\nonumber\\
\frac{1}{R}\, K_{(p+1)i} &\rightarrow\quad e_i &=\quad \sin t^- \partial_{x^i} -\half\cos t^- x_i
\partial_+\nonumber\\
\frac{i}{R^2}\, J_{1q} &\rightarrow\quad e_+ &=\quad \half\partial_+\nonumber\\
\frac{i}{R}\, J_{1i} &\rightarrow\quad f_i^\ast &=\quad \cos t^- \partial_{y^i} +\half\sin t^- y_i
\partial_+\nonumber\\
\frac{i}{R}\, J_{iq} &\rightarrow\quad f_i &=\quad \sin t^- \partial_{y^i} -\half\cos t^- y_i
\partial_+
\eea
where $x^i = r\Omega^i$ and $y^i = u\tilde\Omega^i$ parametrize flat Euclidean spaces.
Together with the angular momenta $K_{ij}$ and $J_{ij}$, they generate the isometries in the Cahen-Wallach space. The broken generators form a Heisenberg algebra
\be\label{eqheis} [\, e_i\;,\; e_j^\ast \, ] = [\, f_i\;,\; f_j^\ast \, ] = \delta_{ij}\,
e_+\ee
with central charge $e_+$.
There is one more Killing vector in this space,
\be \frac{i}{2} (K_{0(p+1)} - J_{1q}) \rightarrow e_- = \partial_-\ee
The ground-state wavefunction
\be \Psi_0 = \Psi_0^{\mathrm{(AdS)}}\times \Psi_0^{(S)} = e^{i(\Delta t-J\phi)}
\cosh^{-\Delta}\rho \cos^J\theta\ee
takes the form
\be\label{eqpsio} \Psi_0 \sim e^{-i(p_+t^++p_-t^-)} e^{-\half p_+ ( r^2+u^2)} \ee
where $\Delta = \half (-p_-+p_+R^2)$, $J = \half (p_-+p_+R^2)$ (this notation differs slightly from~\cite{bib8a}).
The other states are created by repeatedly acting on $\Psi_0$ with
the creation operators $E_I^+$, $F_i^+$, where
\be\label{eqd2} E_i^\pm \equiv e_i \pm ie_i^\ast = e^{\pm it^-} (\partial_{x^i} \pm \halfi x_i\partial_+)\quad,\quad F_i^\pm \equiv f_i \pm if_i^\ast = e^{\pm it^-} (\partial_{y^i} \pm \halfi y_i\partial_+)\ee
The operator $E_i^\pm$
is the Penrose limit of the ladder operator $K_{\pm i}$ (similarly for $F_i^\pm$).
Notice that this action changes the dependence of the wavefunction on $t^-$
(shifting $p_-$) but leaves the quantum number $p_+$
unchanged. Thus these states span the eigenspace of $i\partial_+$ with
eigenvalue $p_+$,
\be\label{eqco5} i\partial_+ \Psi = p_+\Psi\ee
The quadratic Casimir becomes
\bea \mathcal{C} = \lim_{R\to\infty} \frac{1}{R^2}\,
(\mathcal{C}_\mathrm{AdS} + \mathcal{C}_S) &=& 2e_+e_- + (e_i)^2 + (e_i^\ast)^2
+ (f_i)^2 + (f_i^\ast)^2\nonumber\\
&=&  \partial_+\partial_- + (\vec x^{\, 2} + \vec y^{\, 2})\partial_+^2 + \nabla_x^2 + \nabla_y^2 \eea
which is the Laplacian on Cahen-Wallach space.

To see how this is mapped onto the holographic screen (as $\rho\to\infty$),
it is convenient to rotate the generators of the conformal group
on the boundary of AdS~(\ref{eqcong}),
\be\label{eqrot} \mathcal{K}_{AB} \rightarrow e^{\halfi p_- t^-}\mathcal{K}_{AB} e^{-\halfi p_- t^-}\ee
We obtain from (\ref{eqcong}),
\bea\label{eqcong1} \hat\mathcal{K}_{0(p+1)} &=&
\lim_{R\to\infty} \frac{1}{R^2}\, e^{\halfi p_- t^-}\mathcal{K}_{0(p+1)} e^{-\halfi p_- t^-} = \half p_+\nonumber\\
\hat\mathcal{K}_{+i} &=&
\lim_{R\to\infty} e^{\halfi p_- t^-}\mathcal{K}_{+i} e^{-\halfi p_- t^-} = e^{it^-} \left(
\nabla_i-i\Omega_i\partial_-\right) \nonumber\\
\hat\mathcal{K}_{-i} &=&
\lim_{R\to\infty} \frac{1}{R^2}\, e^{\halfi p_- t^-}\mathcal{K}_{-i} e^{-\halfi p_- t^-} =
-p_+\Omega_i\, e^{-it^-}
\eea
where we used the parametrization $\Delta = \half (-p_-+p_+R^2)$.
These vectors on the boundary of AdS
form a Heisenberg algebra
\footnote{For a derivation of this algebra in a quantum field theoretical model
corresponding to AdS$_5\times S^5$, see~\cite{biboku}.}
\be [\, \hat\mathcal{K}_{-i} \;,\; \hat\mathcal{K}_{+j} \, ] = \delta_{ij} p_+\ee
with central charge $p_+$.
The two sets of generators~(\ref{eqcong2}) and (\ref{eqcong1}) act on different spaces which are obtained as different limits of
AdS space ($\rho\to 0$ for CW space; $\rho\to\infty$ for the holographic screen).
AdS space provides a natural bridge between the two extremes but one would
like to map one space onto the other without invoking the full machinery of AdS
space. This map, if it exists, would establish the existence of a ``holographic
screen'' on CW space.

To tackle this issue, let us restrict attention to a subspace of
the Hilbert space on CW space that can be mapped onto the
Hilbert space on the holographic screen of AdS space in the Penrose limit.
A convenient restriction is provided by the constraint
\be\label{eqcon55} \hat\mathcal{D} \Psi = 0\quad,\quad \hat\mathcal{D} = x^i \frac{\partial}{\partial x^i} - i \partial_- +\halfi x^ix_i \partial_+\ee
which commutes with all creation operators~(\ref{eqd2}) as well as the momentum $i\partial_+$ ({\em cf.}~eq.~(\ref{eqco5})),
\be [ \hat\mathcal{D} \;,\; E_i^+] = [ \hat\mathcal{D} \;,\; F_i^+] = 
[ \hat\mathcal{D} \;,\; i\partial_+ ] = 0\ee
It also annihilates the vacuum state $\Psi_0$ (eq.~(\ref{eqpsio})) for $p_- =0$. Thus, it restricts the Hilbert space to the sector which is spanned by all
states created by a string of $E_i^+, F_i^+$ (eq.~(\ref{eqd2})) acting on the
$p_-=0$ vacuum state $\Psi_0$ (eq.~(\ref{eqpsio})). Other sectors may be easily
obtained by rotating as in~(\ref{eqrot}) which amounts to selecting a different
vacuum state $\Psi_0$ with $p_-\ne 0$.

Once (\ref{eqcon55}) is imposed as a constraint, the system is determined by
data specified on a hypersurface in CW space. This hypersurface ought to
intersect each orbit of $\hat\mathcal{D}$
precisely once. In this sense,
transformations generated by $\hat\mathcal{D}$ are similar to gauge transformations and the choice
of hypersurface amounts to a choice of ``gauge.''
A simple choice (gauge-fixing condition) is the cylinder
\be\label{eqcon5} \hat\chi \equiv x^i x_i - \mathcal{R}^2 = 0\ee
The parameter $\mathcal{R}$ is arbitrary. 
The two constraints~(\ref{eqcon55})
and (\ref{eqcon5}) are {\em second class} constraints necessitating the
replacement of Poisson brackets by Dirac brackets,
\be \{ \mathcal{A}\;,\; \mathcal{B}\}_D = \{ \mathcal{A}\;,\; \mathcal{B}\}_P
-\frac{1}{\{ \hat\chi\;,\; \hat\mathcal{D}\}_P}\; \Big( \{\mathcal{A}\;,\;\hat\chi\}_P\{\hat\mathcal{D}\;,\;\mathcal{B}\}_P
- \{\mathcal{A}\;,\;\hat\mathcal{D}\}_P\{\hat\chi\;,\;\mathcal{B}\}_P\Big)\ee
where
$\{ \hat\chi\;,\; \hat\mathcal{D}\}_P = 2\mathcal{R}^2$.
Equivalently, one may parametrize the coordinates appropriately on the
hypersurface~(\ref{eqcon5}) and compute the modification of the algebra of
isometries on the CW space by expressing the Killing vectors in terms of the
reduced set of coordinates.
The Heisenberg algebra~(\ref{eqheis}) is modified to
\bea\label{eqheis1} [E_i^+ \;,\; E_j^-] &=& i\delta_{ij} \partial_+ + \frac{2}{\mathcal{R}^2}\, x_i \left\{ \left( \delta_{jk} - \frac{x_jx_k}{\mathcal{R}^2}
\right) \partial_k +\frac{i}{\mathcal{R}^2} x_j\partial_- \right\} \nonumber \\
\, [E_i^+ \;,\; E_j^+ ] &=& 0 \nonumber \\
\, [E_i^- \;,\; E_j^- ] &=& \frac{2}{\mathcal{R}^2} e^{-2it^-} (x_i\partial_j
-x_j\partial_i)\eea
In the limit $\mathcal{R}\to\infty$, the modification vanishes and the algebra
(\ref{eqheis1}) reduces to a Heisenberg algebra again.
Therefore, $\mathcal{R}$ plays a role
similar to the AdS parameter $R$ in taking the Penrose limit.
To see how this happens explicitly,
we may solve the ``gauge-fixing'' condition~(\ref{eqcon5}), and parametrize
the hypersurface,
\be x^i = \mathcal{R}\Omega^i \ee
We may then use the parametrization of the coordinates to express the conjugate momenta on the hypersurface,
\be \mathcal{R}\frac{\partial}{\partial x^i} = i \Omega_i (\partial_- - \half \mathcal{R}^2\partial_+) + \nabla_i\ee
where we used the constraint~(\ref{eqcon55}) to write $\mathcal{R}\partial_r = 
i(\partial_- - \half \mathcal{R}^2 \partial_+)$.
The CW Killing vectors~(\ref{eqd2}) may be written as
\be\label{eqcong2a}
E_i^\pm = \frac{1}{\mathcal{R}}\, e^{\pm it^-} \left(i\Omega_i (\partial_- -\half \mathcal{R}^2 \partial_+) + \nabla_i \pm \halfi\mathcal{R}^2\Omega_i
\partial_+\right)
\ee
and $F_i^\pm$ remain unchanged.
Using the explicit form~(\ref{eqcong2a}), one may verify that these vectors
satisfy the modified algebra~(\ref{eqheis1}).
These vectors bare a striking resemblance to the generators of the conformal
group on the holographic screen~(\ref{eqcong}).
In the limit
$\mathcal{R}\to\infty$, they turn into the operators
\bea\label{eqcong2v}
\mathcal{E}_i^+ &=& \lim_{\mathcal{R}\to\infty} \mathcal{R}\, E_i^+ = e^{it^-} (\nabla_i -i\Omega_i\partial_-)
\nonumber\\
\mathcal{E}_i^- &=& \lim_{\mathcal{R}\to\infty} \frac{1}{\mathcal{R}}\,
E_i^- = -i e^{-it^-} \Omega_i\partial_+
\eea
in agreement with the generators $\hat\mathcal{K}_{\pm i}$
of the Heisenberg algebra~(\ref{eqcong1}) on the
boundary of AdS in the scaling limit $R\to\infty$.

To summarize, I have recovered the algebra on the holographic screen in the
scaling limit $R\to\infty$~(\ref{eqcong1}) directly on the CW space, without
making use of the AdS space. This was done by restricting the Hilbert space
using the constraint~(\ref{eqcon55}).
After imposing the constraint, the system was determined by data specified
on a hypersurface which was chosen so as to intersect each orbit of
transformations precisely once. In this sense, the transformations generated
by the constraint~(\ref{eqcon55}) are similar to gauge
transformations (albeit rigid) and the choice of hypersurface can be viewed as a
``gauge-fixing'' condition.
The classical Poisson brackets ought to be replaced by Dirac brackets which are then
promoted to commutators upon quantization.
By adopting the gauge~(\ref{eqcon5}),
a surface was selected in Cahen-Wallach space which in the scaling limit
$\mathcal{R}\to\infty$
could be mapped onto the holographic screen (boundary of AdS) in the Penrose limit.
I showed that the
resulting algebra of the CW Killing vectors~(\ref{eqcong2v}) agrees with
the corresponding algebra~(\ref{eqcong1}) on the holographic screen.
This provides a way of identifying a ``holographic screen'' on CW space, even
though no holographic principle, similar to the one in AdS space, exists for
CW space.




\newpage

\begin{thebibliography}{99}

\bibitem{bib1}
R.~Penrose, in {\em Differential geometry and relativity,} pp 217, Reidel,
Dordrecht (1976).
\bibitem{bibsu}
R.~G\"uven, Phys.~Lett.~{\bf B482} (2000) 255, {\tt hep-th/0005061}.
\bibitem{bibst1}
R.~R.~Metsaev,
Nucl.~Phys.~{\bf B625} (2002) 70, {\tt hep-th/0112044}.
\bibitem{bib4a}
N.~Itzhaki, I.~R.~Klebanov and S.~Mukhi,
JHEP {\bf 0203} (2002) 048, {\tt hep-th/0202153}.
\bibitem{bib4aa}
J.~Gomis and H.~Ooguri, Nucl.~Phys.~{\bf B635} (2002) 106,
{\tt hep-th/0202157}.
\bibitem{bib4ab}
L.~A.~Zayas and J.~Sonnenschein,
JHEP {\bf 0205} (2002) 010,
{\tt hep-th/0202186}.
\bibitem{bib4}
M.~Alishahiha and M.~M.~Sheikh-Jabbari, Phys.~Lett.~{\bf B535} (2002) 328,
{\tt hep-th/0203018}.
\bibitem{bib4e}
N.~Kim, A.~Pankiewicz, S.-J.~Rey and S.~Theisen, Eur.~Phys.~J.~{\bf C25} (2002) 327,
{\tt hep-th/0203080}.
\bibitem{bib4f}
T.~Takayanagi and S.~Terashima, JHEP {\bf 0206} (2002) 036,
{\tt hep-th/0203093}.
\bibitem{bib4g}
C.~S.~Chu and P.~M.~Ho, Nucl.~Phys.~{\bf B636} (2002) 141,
{\tt hep-th/0203186}.
\bibitem{bib4h}
U.~Gursoy, C.~Nunez and M.~Schvellinger, JHEP {\bf 0206} (2002) 015,
{\tt hep-th/0203124}.
\bibitem{bib4i}
E.~Floratos and A.~Kehagias, JHEP {\bf 0207} (2002) 031,
{\tt hep-th/0203134}.
\bibitem{bib4j}
M.~Hatsuda, K.~Kamimura and M.~Sakaguchi, Nucl.~Phys.~{\bf B632} (2002) 114,
{\tt hep-th/0202190}.













\bibitem{bib5}
M.~Cvetic, H.~Lu and C.~N.~Pope,
{\tt hep-th/0203082}.
\bibitem{bib5a}
M.~Cvetic, H.~Lu and C.~N.~Pope, Nucl.~Phys.~{\bf B644} (2002) 65,
{\tt hep-th/0203229}.

\bibitem{bib7}
R.~R.~Metsaev and A.~A.~Tseytlin, Phys.~Rev.~{\bf D65} (2002) 126004,
{\tt hep-th/0202109}.
\bibitem{bib7a}
J.~G.~Russo and A.~A.~Tseytlin,
JHEP {\bf 0204} (2002) 021,
{\tt hep-th/0202179}.

\bibitem{bib2}
D.~Berenstein, J.~Maldacena and H.~Nastase,
JHEP {\bf 0204} (2002) 013,
{\tt hep-th/0202021}.

\bibitem{bib3}
J.~Figueroa-O'Farrill and G.~Papadopoulos,
JHEP {\bf 0108} (2001) 036, {\tt hep-th/0105308}.
\bibitem{bib3a}
M.~Blau, J.~Figueroa-O'Farrill, C.~Hull and G.~Papadopoulos,
Class.~Quant.~Grav.~{\bf 19} (2002) L87,
{\tt hep-th/0201081}.
\bibitem{bib3b}
M.~Blau, J.~Figueroa-O'Farrill, C.~Hull and G.~Papadopoulos,
JHEP {\bf 0201} (2002) 047, {\tt hep-th/0110242}.















\bibitem{bib6}
D.~Berenstein, E.~Gava, J.~Maldacena, K.~S.~Narain and H.~Nastase,
{\tt hep-th/0203249}.

\bibitem{bib9}
O.~Aharony, S.~S.~Gubser, J.~Maldacena, H.~Ooguri and Y.~Oz,
Phys.~Rept.~{\bf 323} (2000) 183, {\tt hep-th/9905111}.

\bibitem{bib8}
S.~R.~Das, C.~Gomez and S.~J.~Rey, Phys.~Rev.~{\bf D66} (2002) 046002,
{\tt hep-th/0203164}.
\bibitem{bib8aa}
E.~Kiritsis and B. Pioline, JHEP {\bf 0208} (2002) 048,
{\tt hep-th/0204004}.

\bibitem{bib8a}
R.~G.~Leigh, K.~Okuyama and M.~Rozali, Phys.~Rev.~{\bf D66} (2002) 046004, {\tt hep-th/0204026}.
\bibitem{bib8b}
D.~Berenstein and H.~Nastase, {\tt hep-th/0205048}.


\bibitem{bibego}
G.~Siopsis, Phys.~Lett.~{\bf B545} (2002) 169, {\tt hep-th/0205302}.
\bibitem{biboku}
K.~Okuyama, JHEP {\bf 0211} (2002) 043, {\tt hep-th/0207067}.
\end{thebibliography}
\end{document}